\documentclass[12pt,preprint]{aastex}

\usepackage{amsmath}

\begin{document}

\title{On the formation of the peculiar low-mass X-ray binary IGR J17480$-$2446 in Terzan 5}
\author{
Long Jiang and Xiang-Dong Li}

\affil{$^{1}$Department of Astronomy, Nanjing University, Nanjing 210093, China}

\affil{$^{2}$Key laboratory of Modern Astronomy and Astrophysics (Nanjing University), Ministry of
Education, Nanjing 210093, China}

\affil{$^{}$lixd@nju.edu.cn}

\begin{abstract}
IGR J17480$-$2446 is an accreting X-ray pulsar in a low-mass X-ray binary
harbored in the Galactic globular cluster Terzan 5.
Compared with other accreting millisecond pulsars, IGR J17480$-$2446 is peculiar in its
low spin frequency (11 Hz), which suggests that it might be a mildly recycled neutron star
at the very early phase of mass transfer. However, this model seems to be in contrast
with the low field strength deduced from the kiloHertz quasi-periodic oscillations
observed in IGR J17480$-$2446. Here we suggest an
alternative interpretation, assuming that the current binary system was formed during an
exchange encounter either between a binary (which contains a recycled neutron star)
and the current donor,
or between a binary and an isolated, recycled neutron star.
In the resulting binary, the spin axis of the neutron
star could be parallel or anti-parallel with the orbital axis.
In the later case, the abnormally
low frequency of IGR J17480$-$2446 may result from the spin-down to spin-up evolution
of the neutron star. We also briefly discuss the possible observational implications of the
pulsar in this scenario.
\end{abstract}


\section{Introduction}

Millisecond pulsars are thought to be old neutron stars having been recycled  in low-mass X-ray binaries (LMXBs) \citep{Alpar1982}. According to this scenario, material is transferred to the neutron
star from its companion when it fills its Roche lobe, spinning up the neutron star
to the period of milliseconds, and reducing the surface magnetic field by
several orders of magnitude, from $\sim 10^{11}-10^{12}$ G to $\sim 10^{8}-10^{9}$ G
\citep[][for a review]{Bhattacharya1991}.

Currently there are  14 accreting millisecond pulsars
\citep{pw12}. The first accreting millisecond pulsar is SAX J1808.4$-$3658, which rotates at a
frequency of 401 Hz \citep{WK1998}. Spectral modeling of the iron line
constrained its magnetic field to be
$\sim3\times10^8$ G at the poles \citep{Cackett2009, Papitto2009}, similar to the value
$\sim2\times10^8$ G derived from the spin-down rate measured between outbursts
\citep{Hartman2008, Hartman2009}. With a spin frequency of 599 Hz, IGR J00291$+$5934
is another source showing both spin-up and spin-down during and between outbursts respectively,
suggesting the magnetic field to be $\sim2\times10^8$ G \citep{Hartman2011}.
Both SAX J1808.4$-$3658
and IGR J00291$+$5934 are composed by an accreting pulsar and a brown dwarf companion.
The 435 Hz accreting pulsar XTE J1751$-$305 has a helium white dwarf companion.
Its magnetic field was derived from its spin-down to be $\sim 4\times10^8$ G,
while its spin-up during the 2005 outburst was strongly affected by timing noise \citep{Papitto2008}. From the aforementioned examples, one sees that the spin frequencies and magnetic fields
reported are consistent with the prediction of the recycling scenario.

A new accreting pulsar IGR J17480$-$2446 was detected with the {\em International Gamma-Ray Astrophysics Laboratory} in 2010 October \citep{Bordas2010}. This system is located in the core of
Terzan 5, one of the densest globular clusters in the Galaxy.
It looks like a typical LMXB, with an orbital period of 21.3 h. The binary mass function was
estimated to be $\sim 0.021275(5) M_{\sun}$, indicating
a companion star of mass larger than $0.4 M_{\sun}$  \citep{Papitto2011}.
The most remarkable
feature of IGR J17480$-$2446 is that its rotating frequency is only 11 Hz \citep{Strohmayer2010}, too slow compared with the known spin frequencies ($\sim 185-600$ Hz ) of accreting
millisecond pulsars.

Estimating the magnetic field of IGR J17480$-$2446 is not straightforward.
Assuming that the inner radius of the accretion disk lies between the neutron star's radius
and the corotation radius when the source shows pulsations, \citet{Papitto2011}
and \citet{Cavecchi2011} evaluated
the magnetic field in the range from $\sim 2\times 10^8$ G to $\sim 2\times 10^{10}$ G.
\citet{Miller2011} used the results of a relativistic iron line fit
to estimate the magnetic field at the poles to be $B\sim 10^9$ G.
\citet{Papitto2012} estimated the magnetic field in the range between
$\sim 5\times 10^9$ G and $\sim 1.5\times 10^{10}$ G from the spin-up
rate during outbursts.
Finally, assuming the kiloHertz quasi-periodic oscillation (kHz QPO) frequency as an
orbital frequency at the inner disk radius, one can get a lower limit of the radius. If the disk is
truncated at the magnetospheric radius, the upper limit of the magnetic field of the
neutron star can be derived.
\citet{Barret2012} detected highly significant QPOs soon after the source had moved
from the atoll state to the Z state at frequencies between 800 and 870 Hz, and  suggested
the surface magnetic field be less than $5\times10^8$ G \citep[see also][]{Altamirano2012}.

The above investigations indicate that there is a possibility that the magnetic field of IGR J17480$-$2446 is similar to other accreting millisecond pulsars. If this is the case,
there is interesting implication for its magnetic field evolution.
It is controversial whether there is long-term evolution of the magnetic fields
of rotation-powered neutron stars. However, magnetic field decay in accreting neutron stars
has been widely accepted, and the mechanisms include accelerated Ohmic decay,
vortex-fluxoid interactions, and magnetic burial or screening \citep[][for a review]{payne2008}.
\citet{Shibazaki1989} proposed a phenomenological form relating magnetic field evolution
with accreted mass $\Delta m$ \citep[see also][]{Romani1990},
\begin{equation}
B=\frac{B_{0}}{1+\Delta m/m_{\ast}},
\label{B-decay}
\end{equation}
where $B_0$ is the initial magnetic field, and $m_{\ast}$ is a constant. By fitting to
observations of LMXBs, \citet{Shibazaki1989} found $m_{\ast}\sim10^{-4}\,M_{\odot}$.
\citet{Heuvel-Bitzaraki1995} showed that there is a remarkable correlation between the
magnetic fields and the orbital periods of binary radio pulsars with nearly circular orbits
and low-mass helium white dwarf companions. This relation is consistent with increasing decay
of neutron star magnetic field with increasing amount of matter accreted:
neutron stars with magnetic fields below a few $10^9$ G have accreted material of
$(0.5f)\,M_{\odot}$, where $f\sim 0.5-1$ is the accretion efficiency. From the measured
masses of neutron stars in binary systems, \citet{Zhang2011} also
found that the average mass of millisecond radio pulsars is indeed $\sim 0.2\,M_{\odot}$
heavier that that of other long-period pulsars.

The abnormally low rotation frequency suggests that IGR J17480$-$2446
could be exactly in the process of
becoming an accreting millisecond pulsar. Indeed, observations show that it is
spinning-up at a rate
$\dot{\nu}\approx1.4\times10^{-12}$ Hz\,s$^{-1}$
\citep{Cavecchi2011,Patruno2012}. \citet{Patruno2012} proposed that IGR J17480$-$2446 is a
mildly recycled pulsar which has started a spin-up phase lasting less than a few
$10^5$ yr. This means that IGR J17480$-$2446
is in an exceptionally early RLOF phase.
A potential problem of this scenario is that the incipient
RLOF mass transfer may cause little field reduction according to Eq.~(1)
(see discussion in Section 3).
To account for this, \citet{Patruno2012} assumed that the neutron star underwent
two phase of evolution, i.e., the magneto-dipole spin-down
and the wind accretion spin-down before the current RLOF spin-up phase.
During the wind accretion phase the neutron star
magnetic field $B$ decayed to be  $\sim 10^{10}$ G,
in proportion to the rotation rate, due to the flux-line
vortex line coupling \citep{sri90}. However, this model of magnetic field decay
seems not to be compatible with
observations. For example, the symbiotic X-ray pulsar GX 1$+$4
is believed to possess very
strong magnetic field $B\sim3\times10^{13}$ G with very low spin frequency
$6.3\times10^{-3}$ Hz
\citep{Cui1997}. Other long-period X-ray pulsars such as 4U 2206$+$54 \citep{finger10, reig12},
GX301$-$2 \citep{dor10} and SXP 1062 \citep{fu12} are even though to be accreting magnetars
with $B>10^{14}$ G.

Alternatively, if the magnetic field of IGR J17480$-$2446 is $\sim 10^8-10^9$ G,
Eq.~(1) implies that
it should have accreted a sufficient amount of mass (at least $0.1\,M_{\odot}$),
and it will be difficult to
explain its slow spin in the traditional recycling scenario.
Considering the fact that
Terzan 5 is one of the densest and metal-richest clusters
in our Galaxy \citep{Cohn2002,Ortolani2007}, with 35 rotation-powered millisecond pulsars
discovered so far \citep{Ransom2005, Hessels2006, Pooley2010}, we suggest that the companion star of IGR J17480$-$2446 is not
the original one in its primordial binary, and that the neutron star has undergone close encounter
during which the primordial binary system broke up and formed a triple system.
In the end of the short-interval triple phase, the neutron star captured the current companion and lost its first donor, which had spun it up to
the spin of  milliseconds (along with the magnetic field decayed to $\sim 10^8-10^9$ G).
When the second mass transfer occurred, if the spin angular momentum of
the neutron star was reversed to the orbital angular
momentum of the current companion, the neutron star was spin-down first.
This phase lasted $\sim 10^8$ yr until the spin angular momentum
reduced to zero and succeeded by the current spin-up.
Since the second spin-up epoch started just recently,
it is not abnormal to detect the system with slow spin.

The structure of this work is as follows. In the following section we briefly
review the exchange encounter processes in the globular cluster,
and estimate the formation rate of neutron stars that have evolved from the
 reversed-to-parallel accretion channel.
In Section 3 we describe the possible evolution of IGR J17480$-$2446
in some detail. The observational implications are discussed in Section 4.

\section{The chance of exchange encounters in Terzan 5}

Terzan 5 is reported as the densest globular cluster with a central mass density
$\sim (1 - 4) \times 10^6\,M_{\odot}$pc$^{-3}$ \citep{Lanzoni2010}.
It is composed by two different populations of stars with sub-solar metallicity ($Y=0.26$
and $Z=0.01$)
and an age of $12 (\pm1)$ Gyr, and with supra-solar metallicity ($Y=0.29$ and $Z=0.03$) and
an age of $6 (\pm2)$ Gyr.  Given the high density and old ages of Terzan 5, its X-ray binaries are
likely to be formed during close encounter processes: a neutron star captured tidally during
a close encounter with a single star or took the place of one member of a binary star in an
exchange encounter. In our case we assume that the current donor of IGR J17480$-$2446
has exchanged its original companion in the progenitor binary. We set $m_1$
as the mass of the first donor, which has lost most of its matter and was ejected after the encounter,
$m_2$ as the mass of the neutron star, and $m_3$
as the mass of the incoming object, assumed to be a main sequence star (all the masses are
in the units of $M_{\odot}$).
Following \citet{Heggie1996} we write the semi-analytical exchanging cross section
as follows,
\begin{equation}
\sigma_{\rm ex}=(1.39\times10^3 \,{\rm AU^2})\bar{a} v_{\infty}^{-2}m_{3}^{7/2} M_{123}^{1/6} M_{23}^{1/6} M_{13}^{-5/2} M_{12}^{-1/3} e^k,
\end{equation}
where\\
$M_{12}=m_1+m_2$, $M_{13}=m_1+m_3$,\\
$M_{23}=m_2+m_3$, $M_{123}=m_1+m_2+m_3$,\\
$k=3.70 + 7.49\mu_1 - 1.89\mu_2 - 15.49\mu_1^2 - 2.93\mu_1\mu_2 - 2.92\mu_2^2
+ 3.07\mu_1^3 + 13.15\mu_1^2\mu_2
- 5.23\mu_1\mu_2^2 + 3.12\mu_2^3$,\\
$\mu_1=m_1/M_{12}$, $\mu_2=m_3/M_{123}$.\\
Here $\bar{a}$ is the averaged orbital separation of the binary in units of AU,
and $v_{\infty}$ is the velocity dispersion of the cluster in the units of km\,s$^{-1}$.
This exchange process results in a binary system consisting of a main-sequence
companion and a recycled neutron
star\footnote{Alternatively, an isolated millisecond pulsar may capture a main-sequence
star, or exchange with a less massive star in a normal binary,  to form a LMXB.
}.
The encounter rate is roughly
\begin{equation}
\begin{array}{ll}
\Gamma & \simeq  n_{\rm t} n_{\rm bin} \sigma_{ex} v_{\infty} \nonumber \\
 & \simeq (3.35\times10^{-14} \, {\rm pc^{-3}yr^{-1}})\bar{a} n_{\rm t} n_{\rm bin} v_{\infty}^{-1} m_{3}^{7/2} M_{123}^{1/6} M_{23}^{1/6} M_{13}^{-5/2} M_{12}^{-1/3} e^k,
\end{array}
\end{equation}
where  $n_{\rm t}$ and $n_{\rm bin}$ are the number densities
(in units of $\rm pc^{-3}$) of the target stars and the original binaries, respectively.

We first discuss the number density of the target objects. We
assume that all the stars in the globular cluster were formed more or less simultaneously,
and the initial mass distribution is given by a power law function:
 ${\rm d}N=C_0m^{-1-x}{\rm d}m$ \citep{Salpeter1955,Verbunt1988},
where both the normalization constant $C_0$ and the power index $x$ need to be
derived from the observational data. For Terzan 5 we set the value of $x$ to be in the range of $1-2$
\citep[cf.][]{Verbunt1987, Verbunt1988}.
The normalization constant $C_0$ is dependent on the total stellar mass in the globular cluster,
or the mean mass density $\rho$.
The number density of the target stars ($n_{\rm t}$) can be derived to be
\begin{equation}
\emph{$n_{\rm t}$}=\frac{\int^{m_{\rm up}}_{m_{\rm low}}\rho m^{-1-x}dm}
{\int^{m_{\rm max}}_{m_{\rm min}}m^{-x}dm},
\label{property-1}
\end{equation}
where ${m_{\rm up}}$ and ${m_{\rm low}}$ are the upper and lower mass limits of the target stars
respectively, while ${m_{\rm max}}$ and ${m_{\rm min}}$ are for all stars in the globular cluster.
We take the turnoff mass as the upper limit of
the stellar mass in the cluster, which
can be calculate from the main sequence lifetime of a star with mass $m$ \citep{binary},
\begin{equation}
t_{\rm MS}({\rm Myr})=\frac{m^7+146m^{5.5}+2740m^4+1532}{0.3432m^7+0.0397m^2}
\end{equation}
for $0.25\leq m\leq50$.

Setting $m_{\rm min}=0.1$ and $m_{\rm max}=1.2$ (i.e., the turnoff mass of stars with age of 6 Gyr),
with the reported central mass density of Terzan 5 $\sim (1 - 4) \times 10^6\,M_{\odot}$ pc$^{-3}$
\citep{Lanzoni2010}, we calculate the number density
of stars of mass $0.5-1.2\,M_\odot$\footnote{We assume that the second donor has transferred
at least $\sim 0.1\,M_{\odot}$ to the neutron star (see Eqs. [9] and [10]), so the lower
limit of its initial mass is taken to be $\sim 0.5\,M_{\odot}$.}
to be $n_{\rm t}\sim (1.8-7.2)\times10^5$ pc$^{-3}$
for $x=2$, and $\sim (4.7-18.8)\times10^5$ pc$^{-3}$ for $x=1$. So in the following  we take
$n_{\rm t}\sim 5\times 10^5$ pc$^{-3}$ as a rough estimate.
The number density of the binaries $n_{\rm bin}$ can be estimated by using the total number
of binaries with millisecond pulsars ($N_{\rm b}$) divided by the volume ($V$) of the
cluster core.

The number of binary systems which have undergone exchange encounters with a
reversed-spinning neutron star is
\begin{equation}
N\sim \frac{1}{2}\Gamma T_{\rm p}V
\sim 2.5\times 10 ^{-4}n_{\rm t} a N_{\rm b}f(m)v_\infty^{-1},
\label{property-1}
\end{equation}
where $f(m)=m_{3}^{7/2} M_{123}^{1/6} M_{23}^{1/6} M_{13}^{-5/2} M_{12}^{-1/3}e^k$, and
$T_{\rm p}$ is the time interval between the formation of the original binary
and the encounter, which can be roughly set as $\sim3\times10^9$ yr, the half age of
the metal-rich population in Terzan 5.
A factor of $1/2$ is added to account for the fact that the orbit angular momentum
of the later binary can be either parallel or anti-parallel with the spin of the
neutron star.

Taking typical values for the parameters in Eq.~(5), i.e.,
$m_1\simeq 0.3$, $m_2\simeq 1.4$, $m_3\simeq 0.8$,
$v_{\infty}\sim10$, and $\bar{a}\sim0.02$, we obtain
$N\sim0.5 N_{\rm b}$, suggesting that a considerable fraction of the millisecond binary
pulsars in Terzan 5 may have experienced the specified dynamical interaction.
There are 35 known millisecond pulsars \citep{Ransom2005, Hessels2006, Pooley2010}
in this globular cluster, and the total number of millisecond pulsars may be
$\sim150$ \citep{Bagchi2011}. It is not surprising that all the binary pulsars may
have been formed by dynamical interactions, and probably
half of them might have experienced exchange encounters that leave a reversed-spinning
neutron star in the new binary.

We may expect that  other globular cluster also harbour systems like IGR J17480$-$2446.
For example, in the globular cluster NGC 6440, the central density is
$\sim5\times10^5$ M$_{\odot}$pc$^{-3}$ \citep{Webbink1985}, so
around $5\%$ of the millisecond pulsars might have undergone the exchange evolution.

\section{Spin evolution of the neutron star}

In the last section we argue that the abnormality of IGR J17480$-$2446 may be explained
by assuming that the current donor is not the member of the original binary
but a captured object during a close encounter. In the following we will discuss the
spin evolution of the neutron star in detail.

Due to accretion in the original binary, the neutron star's spin may reach
the equilibrium period \citep{Bhattacharya1991},
\begin{equation}
\emph{$P_{\rm eq}$}\simeq 0.6B_8^{6/7}m^{-5/7} \dot{m}_{17}^{-3/7}R_6^{18/7}\,{\rm ms},
\label{Peq}
\end{equation}
where $B_8$ is the neutron star magnetic field in units of $10^8$ G,
$\dot{m}_{17}$ is the accretion rate in units of $10^{17}$ gs$^{-1}$,
and $R_6$ is the neutron star radius in units of $10^6$ cm.

It is known that, after the neuron star has accreted
$\sim 0.1\,M_\odot$ and its magnetic field has decreased to be $\lesssim 10^9$ G,
its spin period will be insensitive to its initial value \citep[e.g.,][]{Wang2011}.
The required accretion time to reach the equilibrium period is \citep{Alpar1982},
\begin{equation}
t_1\simeq 3.2\times10^7 I_{45}P_{3}^{-4/3}m^{-2/3}
\dot{m}_{17}^{-1}\,{\rm yr},
\label{w-s-1}
\end{equation}
where $P_{3}$ is the spin period in units of 3 ms,
and $I_{45}$ is the momentum of inertia in
units of $10^{45}$ g\,cm$^2$.
This time is considerably shorter than the total duration of mass transfer
in LMXBs (usually $\gtrsim$ a few $10^8$ yr), so it is
likely that when the exchange encounter occurred, the neutron star had already
been spun up to be a millisecond pulsar.


After the capture of the new companion star and the formation of the current binary,
the neutron star first spun down due to magnetic
dipole radiation, but this would not change its spin period significantly, since the evolutionary
time for a rotation-powered millisecond pulsar is usually $\gtrsim$ a few $10^9$ yr.
When the captured star started to fill its RL due to stellar expansion or due to shrinking of the RL
caused by the loss of orbital angular momentum, the neutron star would experience the
second mass transfer that further altered its spin evolution.

If the orbital angular momentum and the neutron star's spin angular momentum were
parallel, accretion onto the neutron star would change its spin period to a new equilibrium
period determined by the current mass accretion rate. However, since $P_{\rm eq}$ is weakly
dependent on the accretion rate as seen in Eq.~(6), the values of the equilibrium periods
should be close to each other\footnote{We assume that the neutron star magnetic field does
not decay further when it has reached the so-called bottom field $\sim 10^8-10^9$ G
\citep[e.g.][]{Zhang2006}.}.
If the orbital and spin angular momenta were
anti-parallel, the subsequent evolution was composed of a spin-down phase followed by
a spin-up phase.
The magnitude of the period derivation in both phases can be described as
\citep{Bhattacharya1991} \begin{equation}
|\dot{P}|\simeq 4.6\times10^{-6}I_{45}^{-1}B_8^{2/7}m^{3/7}R_6^{6/7}
P^2\dot{m}_{17}^{6/7}\,{\rm syr}^{-1}.
\label{p-dot}
\end{equation}
This gives the spin-down time,
\begin{equation}
t_{\rm down}\simeq 7.3\times10^{7}I_{45}B_8^{-2/7}R_6^{-6/7}
P_3^{-1}m^{-3/7}\dot{m}_{17}^{-6/7}\,{\rm yr},
\end{equation}
and the spin-up time to the current period,
\begin{equation}
t_{\rm up}\simeq 2.3\times10^{6}I_{45}B_8^{-2/7}R_6^{-6/7}
P_{100}^{-1}m^{-3/7}\dot{m}_{17}^{-6/7}\,{\rm yr},
\end{equation}
where $P_{100}=P/100$ ms.
The real value of $t_{\rm up}$ could be even larger, since IGR J17480$-$2446 currently appears
as a transient source \citep{Papitto2011,Patruno2012}.
The  typical evolutionary
lifetime of a LMXB is roughly $t_{\rm ev}\sim \Delta m/\dot{m}
\sim 3\times 10^8\dot{m}_{17}^{-1}$ yr for an average accreted mass of $\sim 0.3\,M_{\odot}$.
The observed number of the IGR J17480$-$2446-like systems can then be roughly estimated
as
\begin{equation}
N_{\rm obs}=\frac{t_{\rm up}}{t_{\rm ev}}N\sim (0.007N)I_{45}B_8^{-2/7}R_6^{-6/7}
P_{100}^{-1}m^{-3/7}\dot{m}_{17}^{1/7},
\end{equation}
or $N_{\rm obs}\sim 0.6$ for $N\sim 0.5N_{\rm b}\sim 75$, which suggests there could be
at most one such system in Terzan 5. Obviously the rarity of IGR J17480$-$2446 originates from
its very short duration of the current spin-up phase,
and it will become a millisecond pulsar
again a few $10^7$ yr later.

It is also noted that, according to Eq.~(10), the accreted mass to accomplish the
spin-up to 11 Hz is $\sim 0.002 M_{\sun}$. With this amount of mass, Eq.~(1)
suggests that the magnetic field would have decayed only from $\sim 10^{12}$ G to
$\sim 5\times 10^{10}$ G if this were the first phase of mass accretion.
Specifically, there was enough accreted matter to spin up the neutron star, but it would be
insufficient to substantially reduce the magnetic field to $\sim 10^8-10^9$ G.
In our proposed scenario, this problem does not appear since the neutron star
had already been recycled before the exchange encounter.

\section{Discussion and conclusions}
The newly discovered accreting millisecond pulsar IGR J17480$-$2446 in the globular cluster
Terzan 5 has surprisingly low spin frequency, and has been suggested to be
a mildly recycled pulsar that started a spin-up phase in an exceptionally recent time.
Here we propose an alternative explanation if the magnetic field of IGR J17480$-$2446
is as low as other accreting millisecond pulsars, taking into account
the dense environment of IGR J17480$-$2446.
In dense globular clusters, when a single star interacts with a binary (the neutron star
could either be a member of the binary or the single object), the most probable result
is that one of the binary components is replaced by the single star if
it is the lightest one \citep{krolik84}. In our case
the resulting binary will be composed by a recycled neutron star and a relatively massive
companion star.
The high density of the globular cluster Terzan 5 supports
the possibility of such triple-object close encounter.
The low spin frequency of IGR J17480$-$2446 may be explained as the result of reversed-to-parallel
evolution of the neutron star's spin.

In \citet{Patruno2012}, the system is assumed to be in its incipient mass transfer process,
while in this work, since the donor has transferred some more material through RLOF,
the mass of the companion may be $\sim 0.1-0.2\,M_\odot$ less than its initial mass.
Accordingly, the neutron star has experienced
twice accretion phases, so its mass may be considerably higher than its initial value.
However, it seems difficult to distinguish  our model and
\citet{Patruno2012} in these respects, since detail calculations \citep[e.g.][]{lin2011} show that
the evolutions of LMXBs are rather complicated, depending on the initial masses
of the component stars, the initial orbital periods, and the processes of mass and
orbital angular momentum transfer and loss.
The neutron star magnetic field may serve as a distinct feature.
We assume that the neutron star has experienced long time accretion, so its magnetic filed
has reached the bottom field, $\sim 10^8-10^9$ G, considerably lower than the expected
value of \citet{Patruno2012}. Both the spin evolution and the kHz QPO frequencies
can present constraints on the magnitude of the magnetic field, if the mass accretion rate
of the neutron star can be accurately determined.

Finally it is pointed out that this work is based on the specified relation between
the magnetic field and
accreted mass described by Eq.~(1). As we know that the mechanism for accretion-induced field decay is still uncertain and there may be other forms. For example, in \citet{kiel08}, it is
assumed that the magnetic field decays exponentially with the amount of mass accreted:
\begin{equation}
B=B_0\exp{(-k\Delta M/M_{\sun})},
\end{equation}
where $k$ is a scaling parameter that determines the rate of decay.
For choices of $k = 3000$ and 10000, as suggested by \citet{kiel08}, an accretion of
only $\sim 0.002 M_{\sun}$  can decrease the magnetic field to $\sim 2 \times 10^9$ G
or $< 10^8$ G. Thus, with Eq.~(12) the small amount of accretion
required to spin up the neutron star to 11 Hz would also be sufficient to
highly suppress the magnetic field.
If it can someday be established, via other means, that the capture by the neutron star
of a second companion is necessary, or possibly not needed for IGR J17480$-$2446,
then this might point toward either Eq.~(1) or (12) as the more valid expression
for the field decay in accreting neutron stars.

\begin{acknowledgements}
We are grateful to an anonymous referee for helpful comments.
This work was supported by the Natural Science Foundation of China
under grant number 11133001 and the Ministry of Science, the
National Basic Research Program of China (973 Program 2009CB824800),
and the Qinglan project of Jiangsu Province.

\end{acknowledgements}

\clearpage

\label{lastpage}

\begin{thebibliography}{99}
\expandafter\ifx\csname natexlab\endcsname\relax\def\natexlab#1{#1}\fi


\bibitem[{{Aplar} et~al. (1982)}]{Alpar1982}
Alpar, M. A., Cheng, A. F., Ruderman, M. A. \& Shaham, J. 1982, \nat, 300, 728

\bibitem[{{Altamirano} et~al.(2012)}]{Altamirano2012}
Altamirano, D., Ingram, A., van der Klis, M., et al. 2012, arXiv:1210.1494

\bibitem[{{Bagchi} et~al. (2011)}]{Bagchi2011}
Bagchi, M., Lorimer, D. R. \& Chennamangalam, J. 2011, \mnras, 418, 477

\bibitem[{Barret} (2012)]{Barret2012}
Barret, D. 2012, \apj, 753, 84

\bibitem[{{Bhattacharya \& van den Heuvel} (1991)}]{Bhattacharya1991}
Bhattacharya, D. \& van den Heuvel, E. P. J. 1991, Phys. Rep., 203, 1

\bibitem[{{Bordas} et~al. (2010)}]{Bordas2010}
Bordas, P., Kuulkers, E., Alfonso-Garz\'{o}n, J., et al. 2010, ATel, 2919, 1

\bibitem[{{Boyles} et~al. (2011)}]{boyles11}
Boyles, J., Lorimer, D. R., Turk, P. J., et al. 2011, \apj, 742, 51

\bibitem[{{Cackett} et~al. (2009)}]{Cackett2009}
Cackett, E. M., Altamirano, D., Patruno, A., et al. 2009, \apj, 694, L21

\bibitem[{{Cavecchi} et~al. (2011)}]{Cavecchi2011}
Cavecchi, Y., Patruno, A., Haskell, B., et al. 2011, \apj, 740, L8

\bibitem[{{Cohn} et~al. (2002)}]{Cohn2002}
Cohn, H., Lugger, P. M., Grindlay, J. E. \& Edmonds, P. D. 2002, \apj, 571, 818

\bibitem[{Cui (1997)}]{Cui1997}
Cui, W. 1997, \apj, 482, L163

\bibitem[{{Degenaar \& Wijnands}(2011)}]{Degenaar2011}
Degenaar, N. \& Wijnands, R. 2011, \mnras, 414, L50

\bibitem[{{Doroshenko} et~al.(2010)}]{dor10}
Doroshenko, V., Santangelo, A., Suleimanov, V., et al. 2010, \aap, 515, 10

\bibitem[{{Eggleton}(2006)}]{binary}Eggleton, P. 2006, Evolutionary Processes in
binary and multiple stars, Cambridge University Press,  p.36

\bibitem[{{Finger} et~al.(2010)}]{finger10}
Finger, M. H., Ikhsanov, N. R., Wilson-Hodge, C. A. \& Patel, S. K. 2010, \apj,
709, 1249

\bibitem[{{Fu \& Li} (2012)}]{fu12}
Fu, L. \& Li, X.-D. 2012, \apj, 757, 171


\bibitem[{{Hartman} et~al.(2008)}]{Hartman2008}
Hartman, J. M., Patruno, A., Chakrabarty, D., et al. 2008, \apj, 675, 1468

\bibitem[{{Hartman} et~al.(2009)}]{Hartman2009}
Hartman, J. M., Patruno, A., Chakrabarty, D., et al. 2009, \apj, 702, 1673

\bibitem[{{Hartman} et~al.(2011)}]{Hartman2011}
Hartman, J. M., Galloway, D. K. \& Chakrabarty, D.  2011, \apj, 726, 26

\bibitem[{{Heggie} et~al. (1996)}]{Heggie1996}
Heggie, D. C., Hut, P. \& McMillan, S. L. W. 1996, \apj, 467, 359

\bibitem[{{Hessels} et~al(2006)}]{Hessels2006}
Hessels, J. W. T., Ransom, S. M., Stairs, I. H., et al. Science, 2006, 311, 1901

\bibitem[{{Kiel} et al.(2008)}]{kiel08}
Kiel, P. D., Hurley, J., Bailes, M., \& Murray, J. R. 2008 \mnras, 388, 393

\bibitem[{{Krolik} et al.(1984)}]{krolik84}
Krolik, J. H., Meiksin, A. \& Joss, P. C. 1984, \apj, 282, 466

\bibitem[{{Lanzoni} et~al. (2010)}]{Lanzoni2010}
Lanzoni, B., Ferraro, F. R., Dalessandro, E., et al. 2010, \apj, 717, 653


\bibitem[{{Lin et al.}(2011)}]{lin2011}
Lin, J., Rappaport, S., Podsiadlowski, Ph., Nelson, L., Paxton, B., \& Todorov, P.
2011, \apj, 732, 70

\bibitem[{{Miller} et~al. (2011)}]{Miller2011}
Miller, J. M., Maitra, D., Cackett, E. M., et al. 2011, \apj, 731, L7

\bibitem[{{Ortolani} et~al. (2007)}]{Ortolani2007}
Ortolani1, S., Barbuy, B., Bica, E., et al. 2007, \aap, 470, 1043

\bibitem[{{Papitto} et~ al. (2011)}]{Papitto2011}
Papitto, A., D'A\`{\i}, A., Motta, S., et al. 2011, \aap, 526, L3

\bibitem[{{Papitto} et~ al. (2008)}]{Papitto2008}
Papitto, A., Menna, M. T., Burderi, L., et al. 2008, \mnras, 383, 411

\bibitem[{{Papitto} et~ al. (2009)}]{Papitto2009}
Papitto, A., Di Salvo, T.,  D'A\`{\i}, A., et al. 2009, \aap, 493, L39

\bibitem[{{Papitto} et~ al. (2012)}]{Papitto2012}
Papitto, A., Di Salvo, T., Burderi, L., et al. 2012, \mnras, 423, 1178

\bibitem[{{Patruno} et~al. (2012)}]{Patruno2012}
Patruno, A., Alpar, M. A., van der Klis, M. \& van den Heuvel, E. P. J. 2012, \apj, 752, 33

\bibitem[{{Patruno \& Watts}(2012)}]{pw12}
Patruno, A. \& Watts, A. L. 2012, to appear in ``Timing neutron stars: pulsations,
oscillations and explosions", T. Belloni, M. Mendez, C.M. Zhang Eds., ASSL,
Springer (arXiv:1206.2727)

\bibitem[{{Payne et al.}(2008)}]{payne2008}
Payne, D. J. B., Vigelius, M., \& Melatos, A. 2008, in A decade of accreting millisecond
X-ray pulsars. AIP Conference Proceedings, Vol. 1068, p. 144

\bibitem[{{Pooley} et~al. (2010)}]{Pooley2010}
Pooley, D., Homan, J., Heinke, C., et al. 2010, ATel, 2974, 1


\bibitem[{{Ransom} et~al. (2005)}]{Ransom2005}
Ransom, S. M., Hessels, J. W. T., Stairs, I. H., et al. 2005, Science, 307, 892

\bibitem[{{Reig} et~al. (2012)}]{reig12}
Reig, P., Torrej\'on, J. M. \& Blay, P. 2012, \mnras, 425, 595

\bibitem[{{Romani}(1990)}]{Romani1990}
Romani R. W. 1990, \nat, 347, 741

\bibitem[{{Salpeter} (1955)}]{Salpeter1955}
Salpeter, E. E. 1955, \apj, 121, 161s


\bibitem[{{Shibazaki} et~al. (1989)}]{Shibazaki1989}
Shibazaki N., Murakami T., Shaham J., \& Nomoto K., 1989, \nat, 342, 656

\bibitem[{{Srinivasan} et~al. (1990)}]{sri90}
Srinivasan, G., Bhattacharya, D., Muslimov, A. G. \& Tsygan, A. J. 1990, Curr. Sci., 59, 31

\bibitem[{Strohmayer \& Markwardt (2010)}]{Strohmayer2010}
Strohmayer, T. E. \& Markwardt, C. B. 2010, ATel, 2929, 1

\bibitem[{van den Heuvel \& Bitzaraki (1995)}]{Heuvel-Bitzaraki1995}
van den Heuvel, E. P. J. \& Bitzaraki, O. 1995, \aap, 297, L41

\bibitem[{{Verbunt}(1988)}]{Verbunt1988}
Verbunt, F. 1988, Adv. Space Res., 8, 529

\bibitem[{{Verbunt \& Hut}(1987)}]{Verbunt1987}
Verbunt, F. \& Hut, P. 1987, in The origin and evolution of nutron stars, IAU Symp.
No.125, eds. Helfand, D. J. \& Huang, J.-H., Reidel, Dordrecht, p.187

\bibitem[{Wang} et~al.(2011)]{Wang2011}
Wang, J., Zhang, C.-M., Zhao, Y.-H., et al. 2011, \aap, 526, A88

\bibitem[{Webbink}(1985)]{Webbink1985}
Webbink, R. F. 1985, in Dynamics of star clusters, IAU Symp. No.113, eds.
Goodman, J. \& Hut, P. Reidel, Dordrecht, p.541

\bibitem[{{Wijnands \& van der Klis}(1998)}]{WK1998}
Wijnands, R. \& van der Klis, M.  1998, \nat, 394, 344

\bibitem[{{Zhang \& Kojima}(2006)}]{Zhang2006}
Zhang, C.-M. \& Kojima, Y. 2006, \mnras, 366, 137

\bibitem[{{Zhang et al.}(2011)}]{Zhang2011}
Zhang, C.-M. Wang, J., Zhao, Y.-H., Yin, H.-X., Song, L.-M. et al. 2011, \aap, 527, A83

\end{thebibliography}
\end{document}